\documentclass{elsarticle}      
\journal{Discrete Applied Mathematics}     
\usepackage[T1]{fontenc}
\usepackage{pdflscape}
\usepackage{multirow}
\usepackage[francais]{babel}     
\usepackage[scale=0.75]{geometry}
\usepackage{tikz}
\usepackage{graphicx}
\usepackage{subfig}
\usepackage{amssymb}
\usepackage{colortbl}
\usepackage{amsmath}
\usepackage[algoruled,vlined,french,linesnumbered]{algorithm2e}
\usepackage{float}
\usepackage{longtable}
\SetKwInOut{Input}{Input}
\SetKwInOut{Output}{Output}
\newcommand{\mscp}{$MSCP$ }
\newcommand{\gcp}{$GCP$ }

\newcommand{\G}{\textit{G}\xspace}
\newcommand{\X}{\mathcal{X}\xspace}

\newcommand{\s}{\textit{s}\xspace}

\newtheorem{exemple}{Example \\}
\newtheorem{definition}{Definition}
\newtheorem{preuve}{Proof}
\newtheorem{prop}{Property}

\author{Cl\'ement Lecat, Corinne Lucet, Chu-Min Li}

\address{Laboratoire de Mod\' elisation, Information et Syst\`eme EA 4290, Universit\' e de Picardie Jules Verne, Amiens, France\\ } 

\ead{clement.lecat, corinne.lucet,chu-min.li@u-picardie.fr}
\begin{document}
\begin{frontmatter}

\title{	Sum Coloring : New upper bounds for the chromatic strength}

%\author{
%	\large{\textbf{Laboratoire : }MIS, \textbf{\'{E}quipe : }GOC}\\\\
%	\large{\textbf{Auteur : }Clément Lecat, Corinne Lucet, Chu-Min Li }
%	}
	
	%\maketitle
	\thispagestyle{empty}
	
%	\newline
	
\begin{abstract}
The Minimum Sum Coloring Problem  ($MSCP$) is derived from the Graph Coloring Problem ($GCP$) by associating a weight to each color. The aim of $MSCP$ is to find a coloring solution of a graph such that the sum of color weights is minimum. $MSCP$ has important applications in fields such as scheduling and VLSI design. We propose in this paper new upper bounds of the chromatic strength, i.e. the minimum number of colors in an optimal solution of $MSCP$, based on an abstraction of all possible colorings of a graph called motif. Experimental results on standard benchmarks show that our new bounds are significantly tighter than the previous bounds in general, allowing to reduce substantially the search space when solving $MSCP$.
\end{abstract}

\end{frontmatter}

\section {Introduction}

The Graph Coloring Problem (\gcp) is an important NP-hard combinatorial problem. A lot of effort has been devoted to study it. Two main types of algorithms (also called solvers) are developed for solving \gcp: exact methods and approximate methods. Exact methods aim at finding an optimal solution of the problem, including the approaches based on branch-and-bound schema \cite{Men06}, on graph decomposition \cite{Lucet06}, and on SAT solving by encoding the problem into an equivalent propositional formula \cite{ZLHX14}. The minimum number of colors needed to color a graph $G$ is called the {\em chromatic number} of $G$ and is denoted by $\chi(G)$. The approximate methods aim at finding an upper bound or a lower bound of the optimal solution of the problem, including: greedy algorithms such as the famous DSATUR \cite{Brelaz79}, and various heuristic or meta-heuristic algorithms \cite{Porumbel,Hertz2008,HAO2010}. In the literature, these methods are usually evaluated on standard benchmarks such as DIMACS and COLOR \cite{dimacs2,color023}.

The Minimum Sum Coloring Problem (\mscp) is derived from \gcp and is introduced in 1989 by Kubicka et Schwenk \cite{CSC-89}, by associating a weight to each color. The aim of \mscp is to find a valid coloring solution that minimizes the sum of color weights. The  minimum number of colors in an optimal solution of \mscp for a graph $G$ is called the chromatic strength of $G$ and is denoted by $s(G)$. Note that $s(G)$ can be bigger than $\chi(G)$. $MSCP$ has important applications in fields such as scheduling, VLSI design and ressource allocation \cite{bar,Malafiejski04}. For example, to calculate the best quality of service in a distributed system with shared resource amounts to solve $MSCP$. The main results on \mscp include the theoretical bounds \cite{jan,Alavi-et-al89,Malafiejski04,bar2} and the structural properties relative to the graph families for which efficient \mscp algorithms exist. Recently, heuristics and meta-heuristics for \mscp are proposed in \cite{LT} and \cite{sgh,JinHH14,una,Mouk2010} respectively, giving bounds for DIMACS and COLOR graphs, and exact methods are proposed in \cite{lecatetalcp15}.

When solving \gcp and $MSCP$ for a graph $G$, an algorithm generally has to explore the search space of \gcp and \mscp that grows exponentially with the number of colors to be considered. In practice it is substantially harder to reduce the number of colors to be considered when solving \mscp than when solving $GCP$. In fact, for $GCP$ when a valid coloring solution with $k$ colors is found, the sub-space with $k$ or more colors can be pruned. However, this is not the case for $MSCP$, because the optimal solution of \mscp can involve more than $k$ colors. So, establishing a tight upper bound of $s(G)$ is essential to solve $MSCP$.

Unfortunately, there are few works in the literature allowing to derive a tight upper bound of $s(G)$. The only two existing upper bounds in our knowledge are proposed in \cite{mitch} and in \cite{hajia} respectively. In this paper, we propose two new upper bounds of $s(G)$ by exploring an abstraction of the set of coloring solutions of $G$ called {\em motif}. The notion of motif was already used by Bonomo and Valencia-Pabon in \cite{Bon1, Bon2} to solve $MSCP$ for P$_4$-sparse graphs.  However, it is the first time in our knowledge that motifs are used for upper bounding $s(G)$. By skillfully identifying and excluding those motifs that cannot correspond to an optimal solution of MSCP, we derive the two new upper bounds of $s(G)$ from the remaining motifs. The experimental results on standard benchmarks DIMACS and COLOR \cite{dimacs2,color023} for coloring problems show that our bounds are substantially better than the existing bound proposed in \cite{hajia} in general. In some instances, the gain is greater than 200 colors. The other existing bound proposed in \cite{mitch} is not compared because it is trivial.

This paper is organized as follows. Section 2 presents the preliminaries necessary for our approach. 
%includes the definitions of GCP, MSCP, as well as two related problems MaxClique and MaxStable, 
%the motif notion and a dominance relation between motifs needed to establish the new upper bounds. 
Section 3 presents our approach for identifying the motifs that cannot correspond to an optimal solution of $MSCP$ and for computing the new upper bounds of $s(G)$. Section 4 compares our new upper bounds with the existing bound proposed in \cite{hajia} on the DIMACS and COLOR graphs. Section 5 concludes.

\section{Preliminaries}
\subsection{Basic definitions, GCP, MSCP, MaxClique and MaxStable}

We consider an undirected graph $G=(V,E)$, where $V$ is a set of vertices ($|V| = n$) and  $E\subseteq V^2$ a set of edges. The set of adjacent (or neighbor) vertices of $v \in V$, denoted by $\mathcal{N}$, is defined as : $\mathcal{N}(v)=\{ u \mid (u,v)\in E \}$. The degree $d(v)$ of a vertex $v$ is the number of its adjacent vertices, i.e., $d(v)= | \mathcal{N}(v) |$. The degree of a graph, denoted by $\Delta(G)$, is  $max\{d(v) \mid v \in V \}$.  A clique $C$ is a subset of $V$ such that $\forall u,v \in C$, $(u,v)\in E$. A stable set $S$ is  a subset of $V$ such that  $\forall u,v \in S$, $(u,v) \not \in E$. The complement graph of $G$ is defined as $\overline G = (V, \overline E)$ where $\overline E = V^2\setminus E$. A clique in $G$ is a stable set in $\overline G$ and vice versa. A graph $G'=(V',E')$ is a subgraph of $G$ induced by $V'$ if $V' \subseteq V $ and $E'=V'^2\cap E$. 

A coloring of a graph $G$ with $k$ colors is a function $c: V \mapsto \{1, 2,  \ldots ,k\}$ that assigns to each vertex  $v\in V$ a color $c(v)$.  A coloring is valid, if $\forall (u,v) \in E$, $c(u) \neq c(v)$. We denote a coloring of  $G$ with $k$ colors by $X=\{X_1,X_2,\ldots,X_k\}$, where $X_i=\{ v \in V \ \mid \  c(v)=i \}$ is called a color class.  The Graph Coloring Problem ($GCP$) consists in finding a valid coloring $X$ of $G$ with minimum $k$. Such $k$ is called the {\em chromatic number} of $G$, and is denoted by $\chi(G)$. $GCP$ is NP-Hard \cite{Gar}.

The Minimum Sum Coloring Problem ($MSCP$) is derived from $GCP$ by associating a  weight $w_i$ with each color $i$.
In this paper, we consider $w_i=i$. We denote by $\Sigma(X)$ the sum of color weights of a coloring:

$$
 \Sigma(X)=1 \times | X_1 | + 2 \times | X_2 | + ... + k \times | X_k |
$$

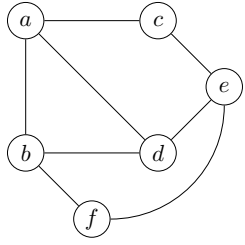
\begin{figure}[!ht]
\centering
\scalebox{0.88}{\begin{tikzpicture}
\tikzstyle{noeud}=[draw, circle, minimum height=0.55cm, minimum
  width=0.55cm, inner sep=0pt]
  
	\node[noeud] (A) at (-2, 2) {$a$};
	\node[noeud] (B) at (-2, 0) {$b$};
	\node[noeud] (C) at (0, 2) {$c$};
	\node[noeud] (D) at (0, 0) {$d$};
	\node[noeud] (E) at (1, 1) {$e$};
	\node[noeud] (F) at (-1, -1) {$f$};

	\draw[-, >=stealth ] (A) -> (C);
	\draw[-, >=stealth ] (A) -> (D);
	\draw[-, >=stealth ] (A) -> (B);
	\draw[-, >=stealth ] (D) -> (B);
	\draw[-, >=stealth ] (F) -> (B);
	\draw[-, >=stealth ] (D) -> (E);
	\draw[-, >=stealth ] (C) -> (E);
	\draw[-, >=stealth ] (F) to[bend right=45] (E);
\end{tikzpicture}}
\caption{ \small A simple graph}
\label{simplegraph}
\end{figure}

\begin{exemple}
\label{defsom}

Refer to the graph in Figure \ref{simplegraph}, $X=\{\{a, e\}_1,  \{b\}_2, \{c, d,f\}_3\}$ is a valid coloring. The  vertices  $a$ and $e$  are colored with the color $1$, the vertex $b$ with the color $2$, and the vertices $c$, $d$ and $f$ with the color $3$. The sum coloring is $\Sigma(X)=1 \times 2 + 2 \times 1 + 3 \times 3 = 13$.
\end{exemple}

Given a graph $G$, $MSCP$ consists in finding a valid coloring $X$ of $G$ with the minimum  sum of color weights $\Sigma(X)$. This minimum sum is called the chromatic sum of  $G$  and is denoted by $\Sigma (G)$:
$$
\Sigma(G)=min \{\Sigma(X)  \mid X~is~a~valid~coloring~of~G \}
$$

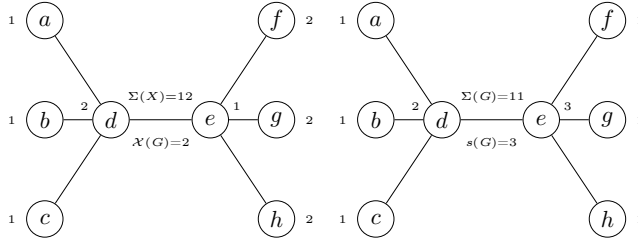
\begin{figure}[h]
\centering
\scalebox{0.88}{\begin{tikzpicture}
\tikzstyle{noeud}=[draw, circle, minimum height=0.55cm, minimum
  width=0.55cm, inner sep=0pt]
  
	\node[noeud] (A) at (-2, 1.5) {$a$};
	\node at (-2.5, 1.5) {\tiny{$1$}};
	\node[noeud] (B) at (-2, 0) {$b$};
	\node at (-2.5, 0) {\tiny{$1$}};
	\node[noeud] (C) at (-2, -1.5) {$c$};
	\node at (-2.5, -1.5) {\tiny{$1$}};
	\node[noeud] (D) at (-1, 0) {$d$};
	\node at (-1.4, 0.2) {\tiny{$2$}};

	\node[noeud] (E) at (0.5, 0) {$e$};
	\node at (0.9, 0.2) {\tiny{$1$}};

	\node[noeud] (F) at (1.5, 1.5) {$f$};
	\node at (2, 1.5) {\tiny{$2$}};
	\node[noeud] (G) at (1.5, 0) {$g$};
	\node at (2, 0) {\tiny{$2$}};
	\node[noeud] (H) at (1.5, -1.5) {$h$};
	\node at (2, -1.5) {\tiny{$2$}};

	\node[noeud] (A1) at (3, 1.5) {$a$};
	\node at (2.5, 1.5) {\tiny{$1$}};
	\node[noeud] (B1) at (3, 0) {$b$};
	\node at (2.5, 0) {\tiny{$1$}};
	\node[noeud] (C1) at (3, -1.5) {$c$};
	\node at (2.5, -1.5) {\tiny{$1$}};
	\node[noeud] (D1) at (4, 0) {$d$};
	\node at (3.6, 0.2) {\tiny{$2$}};
	\node[noeud] (E1) at (5.5, 0) {$e$};
	\node at (5.9, 0.2) {\tiny{$3$}};
	\node[noeud] (F1) at (6.5, 1.5) {$f$};
	\node at (7, 1.5) {\tiny{$1$}};
	\node[noeud] (G1) at (6.5, 0) {$g$};
	\node at (7, 0) {\tiny{$1$}};
	\node[noeud] (H1) at (6.5, -1.5) {$h$};
	\node at (7, -1.5) {\tiny{$1$}};

	\node at (-0.25, 0.35) {\tiny{$\Sigma$($X$)$=$$12$}};	
	\node at (-0.25, -0.35) {\tiny{$\X(\G)$=$2$}};	
	\node at (4.77, 0.35) {\tiny{$\Sigma(\G)$=$11$}};	
	\node at (4.75, -0.35) {\tiny{$\s(\G)$=$3$}};

	\draw[-, >=stealth ] (A) -> (D);
	\draw[-, >=stealth ] (B) -> (D);
	\draw[-, >=stealth ] (C) -> (D);
	\draw[-, >=stealth ] (D) -> (E);
	\draw[-, >=stealth ] (E) -> (F);
	\draw[-, >=stealth ] (E) -> (G);
	\draw[-, >=stealth ] (E) -> (H);
	
	\draw[-, >=stealth ] (A1) -> (D1);
	\draw[-, >=stealth ] (B1) -> (D1);
	\draw[-, >=stealth ] (C1) -> (D1);
	\draw[-, >=stealth ] (D1) -> (E1);
	\draw[-, >=stealth ] (E1) -> (F1);
	\draw[-, >=stealth ] (E1) -> (G1);
	\draw[-, >=stealth ] (E1) -> (H1);

\end{tikzpicture}}
\caption{\small An optimal solution for \gcp: $X$=\{\{$a$, $b$, $c$, $e$\}$_1$, \{$d$, $f$, $g$, $h$\}$_2$\} with sum of color weights equal to 12, and an optimal solution for \mscp: $X'$=\{\{$a$, $b$, $c$, $f$, $g$, $h$\}$_1$, \{$d$\}$_2$, \{$e$\}$_3$\}\} with sum of color weights equal to 11.} 
\label{fig2}
\end{figure}

Kubicka and Schwenk proved that  $MSCP$ is NP-Hard \cite{CSC-89}. An optimal solution of $GCP$ does not necessarily correspond to an optimal solution for $MSCP$. For example, the optimal solution of $GCP$ for the graph in Figure \ref{fig2} uses 2 colors, for which the sum of color weights is 12, while an optimal solution of \mscp for this graph uses 3 colors. The chromatic sum of the graph is 11. The minimum number of colors in an optimal solution of $MSCP$ of a graph $G$ is called {\em chromatic strength} (or simply {\em strength}) of $G$, and is denoted by $s(G)$.

A valid coloring of a graph $G=(V,E)$ with $k$ colors $c: V \mapsto \{1, 2,  \ldots ,k\}$ is to be found in a set of colorings of cardinality $k^{|V|}$, which forms the search space for both $GCP$ and $MSCP$. The number $k$ must be large enough to achieve an optimal solution. While \gcp and \mscp are both NP-hard, \mscp is much harder to solve than \gcp in practice, because it is much more complex to prune search space when solving $MSCP$. So, determining an upper bound of $s(G)$ as tight as possible is essential for solving $MSCP$.

Given a graph $G$, the MaxClique (MaxStable) problem consists in finding a clique (stable set) with the maximum cardinality in $G$. Note that finding a maximum stable set in $G$ is equivalent to finding a maximum clique in $\overline G$.

\subsection{Major coloring}

For a given coloring $X$, each color class $X_i$ is a stable set. We can exchange two colors $i$ and $j$ without impacting the validity of $X$.  The set of colorings that can be achieved by such exchanges from $X$, are symmetric and form an equivalence class, denoted by $\Theta(X)$. All colorings of $\Theta(X)$ use the same number of colors, but they do not give the same sum of color weights. Refer to Figure \ref{simplegraph}, the sum of color weights of the coloring $\{\{a, e\}_1, \{b\}_2, \{c, d,f\}_3\}$ is 13, while the sum of color weights of the coloring $\{\{a, e\}_1, \{c, d,f\}_2, \{b\}_3\}$ is 11. 
 Therefore, we define the notion of {\em major coloring}.

\begin{definition}
A major coloring, denoted by $X^m$,  is a coloring $X^m=\{X_1, X_2, \ldots, X_k\}$ such that  $|X_1|\geq|X_2|\geq \ldots \geq |X_k|$.
\label{xtheta}
\end{definition} 

\begin{exemple}
\label{id}
Refer to Figure \ref{simplegraph}, the coloring $X=\{\{c, d,f\}_1,\{a, e\}_2, \{b\}_3\}$ is a major coloring of the graph.
 $\Sigma(X^m)=10$.
\end{exemple}

%Note that $\Theta(X)$ can contain several different major colorings when several color classes have the same cardinality. 

The following property is a direct consequence of Definition \ref{xtheta}.

\begin{prop}
Let $\Theta(X)$ be the set of symmetric colorings of $X$ and $X^m$ a major coloring of $\Theta(X)$, then $\forall X' \in \Theta(X)$, $\Sigma(X^m)\leq \Sigma(X')$.
\end{prop}

Consequently, only major colorings need to be considered when solving $MSCP$.  In the sequel, all colorings we consider are major and are simply written as $X$.

\subsection{Motifs}
Any major coloring $X$ with $k$ colors corresponds to a non-increasing sequence $p$ of integers: ($|X_1|$, $|X_2$, $\ldots$, $|X_k|$), called the {\em motif} of $X$. The $i^{th}$ integer is denoted by $p[i]=|X_i|$. %Note that $|X_1|\geq|X_2|\geq \ldots \geq |X_k|$.

The sum of color weights of $X$ can be computed as:

\begin{equation}
\label{equaMotif}
\Sigma(X)=\Sigma(p)=1\times p[1] + 2\times p[2] + .... + k\times p[k]
\end{equation}

\begin{exemple}
The  motif corresponding to $X= \{\{c, d,f\}_1, \{a,e\}_2, \{b\}_3 \}$ is $p=(3,2,1)$, with $p[1]=3$, $p[2]=2$ and $p[3]=1$. $\Sigma(X)=\Sigma(p)=10$.
\end{exemple}

The interest of the motif notion is that a motif provides an abstraction of several colorings, that is essential for $MSCP$. For example, refer to Figure \ref{simplegraph}, the two different colorings $\{c, d,f\}_1, \{b,e\}_2, \{a\}_3$ and $  \{a,e,f\}_1, \{b,c\}_2, \{e\}_3$ can be represented by the same motif $p=(3,2,1)$. Two different motifs represent necessarily different colorings. Nevertheless, a motif can represent an invalid coloring, because it does not include any structural property of a graph. For example, refer to Figure 1, the motifs $(5,1)$ and $(4,1,1)$ do not represent any valid coloring of the graph. 
As we will show in Section \ref{sectionUBs}, some characteristics of a graph can be used to exclude a part of motifs representing invalid colorings, so that we can derive interesting properties for $MSCP$ from the remaining motifs.
We denote by $\phi(n)$ the set of all motifs for any graph with $n$ vertices, and $\phi(n,k)$ the set of motifs with $k$ colors for any graph with $n$ vertices.

%Nous pouvons donc résumer notre système de représentation de l'ensemble des solutions de MSCP comme suit : un motif p représente un ensemble de colorations 

$$
\begin{array}{l}
\phi (n,k)=\{ p\in \phi (n) \mid |p|=k\} \\
\phi(n)=\bigcup\limits_{k=1}^n \{\phi (n,k)\}
\end{array}
$$

The number of motifs in $\phi(n)$ is equal to the number of partitions of $n$ into non-increasing integers, which grows exponentially with $n$. Hardy and Ramanujan\cite{HR}  give an approximate number of partitions of $n$ into non-increasing integers :
 
\begin{equation}
|\phi(n)|\sim\frac{1}{4n\sqrt{3}}\ e^{\pi\sqrt{2n/3}}
\end{equation}

Table $1$ illustrates the exponential growth of the cardinality of $\phi(n)$. 

\begin{table}[h]
\label{tabphi}
\begin{tabular}{|c|c|c|c|c|c|c|c|c|c|c|c|c|c|c|}

\hline
n & 1 & 2 & 3 & 4 & 5 & 6 & 7 & 8 & 9 & 10 & 20 & 50 & 100 & 150\\
\hline
$|\phi (n)|$ &  1 & 2 & 3 & 5 & 7 & 11 & 15 & 22 & 30 & 42 & 697 & 204,226 & 190,569,292 & 40,853,235,313\\
\hline
\end{tabular}
\caption{\small Cardinality of $\phi(n)$ . }
\end{table}

Our approach works in $\phi(n)$ to find a tight upper bound of the chromatic strength of any graph with $n$ vertices, using the dominance relation between the motifs defined in the next subsection.

\subsection{Dominance relation}

The dominance relation, denoted by $\succeq$, between the motifs of $\phi(n)$ was introduced by Bonomo and Valencia in \cite{Bon1,Bon2} to compute the chromatic sum of  a subset of $P_4$-sparse graphs, or an upper bound of the chromatic sum for general $P_4$-sparse graphs. We adapt their definition for our approach to derive tight upper bounds of the chromatic strength.

\begin{definition}
Let $p$ and $q$ be two motifs in $\phi(n)$. We say that $p$ dominates $q$, denoted by $p \succeq q$, if and only if $\forall t$ such that $1\leq t \leq min\{ |p|,|q|\}$, $\sum\limits_{x=1}^{t}p[x] \geq \sum\limits_{x=1}^{t}q[x]$. \label{dominance}
\end{definition}

The main difference of Definition \ref{dominance} with the original definition in \cite{Bon1} is that in this paper we explicitly require $\sum\limits_{x=1}^{|p|}p[x]=\sum\limits_{x=1}^{|q|}q[x]=n$, because $p$ and $q$ are both in $\phi (n)$.

%\begin{exemple}
%Let $p$ and $q$ be two motifs, with $p=(5,2,1)$ and $q=(4,2,2)$.\\
%$
%\begin{array}{ll}
%For : &t=1 \Rightarrow 5 \geq 4 \\
%
%&t=2 \Rightarrow 5 + 2 \geq 4 + 2 \\
%&t=3 \Rightarrow 5 + 2 + 1 \geq 4 + 2 + 2 \\
%
%\end{array}
%$\\
%So  $p \succeq q$. 
%\end{exemple}

Two motifs are not necessarily comparable, as shown in Example \ref{comparaison}. 

\begin{exemple}
\label{comparaison}
Let $p$ and $q$ be two motifs of a graph $G$ with  $15$ vertices, $p=(9,3,3)$ and $q=(8,6,1)$. \\
If $t = 1$ : $\sum\limits_{i=1}^{1} p[i] > \sum\limits_{i=1}^{1}q[i]$ ( 9 > 8 ). \\
If $t=2$ : $\sum\limits_{i=1}^{2} p[i] < \sum\limits_{i=1}^{2}q[i]$ ( 9+3 < 8+6 ).\\
$p \not \succeq q$ and $q \not \succeq p$, i.e., $p$ and $q$ are not comparable.
\end{exemple}

So, the dominance relation $\succeq$ is a partial order. It is easy to prove the following property that makes the dominance relation useful for \mscp \cite{Bon1,Bon2}.

%%La notion de coloration majeure permet une représentation sans symétrie de l'ensemble des solutions 
%pour MSCP. En effet, seule la coloration $X^\theta$ de $\Theta(X)$ est considérée. De plus, il est a noter que
%
%\begin{prop}
%The dominance relation $\succeq$  is a partial order.
%\end{prop}

\begin{prop}
Let $G$ be a graph, $p$ and $q$  two motifs corresponding to two valid colorings  $X$ and $X'$ of $G$, respectively. If $p \succeq q$ then $\Sigma(X)\leq \Sigma(X')$.
\end{prop}

%The exact methods and the approximate methods developed for MSCP, explore the search 
%space to find a solution. The size of the search space is depending on the initial upper bound 
%of the strength based on the motifs order. Indeed, let $k$ be the initial colors number, the size 
%of search space for MSCP is $k^{|V|}$, for maxSAT and minSAT $2^{k \times |V|}$... Thus, 
%we have a great interest to have relevant overestimation of it. We propose  an algebraic upper 
%bound and an algorithmic upper bound.

\section{Tight upper bounds for the chromatic strength}
%\subsection{Introdution}

Much effort is spent to establish the relationship between the structural properties of a graph $G$ and its chromatic sum \cite{bar2,sala,jan2,Bon2}. However, less attention is paid to the chromatic strength. Property \ref{trivialUB}  states a trivial upper bound of $s(G)$ found in \cite{mitch} :

\begin{prop}
\label{trivialUB}
$$
s(G)\leq \Delta(G)+1
$$
\end{prop}

A better upper bound based on a valid coloring of $G$ is given in \cite{hajia} and is stated in Property \ref{Haj}. 

\begin{prop}
\label{Haj}
Let $G$ be a graph and $X$ a valid coloring with $k$ colors, then
$$
s(G)\leq\left\lceil\frac{\Delta(G)+\chi(G)}{2}\right\rceil\leq\left\lceil\frac{\Delta(G)+k}{2}\right\rceil
$$

\end{prop}

In this section, we propose two new upper bounds of  $s(G)$ based on a valid coloring of $G$ by exploring $\phi(n)$. For this purpose, we first present a total order in $\phi(n)$ and an algorithm allowing to assign an index to each motif in $\phi(n,k)$, which is needed for the understanding of the new upper bounds. Then we present the bounds after explaining their principle.

\subsection{Establishing a total order in $\phi(n)$}

The set of motifs  $\phi(n)$ can be sorted in the following order: $\phi(n,1),\phi(n,2),\ldots,\phi(n,n)$, then each $\phi(n,k)$ is sorted in the  decreasing  lexicographical order. As an example, Table $2$ lists all motifs in $\phi(8)$ in the above order.

\begin{table}[h]
\label{tab:partition}
\begin{center}
\begin{tabular}{|c|l|c|}
\hline
 $\phi$(n,k)  & \multicolumn{1}{|c|}{p$\in\phi$(n,k)}& $\Sigma(p)$  \\
\hline
$\phi$(8,1) & (8) & 8 \\
\hline
 \multirow{4}{*}{$\phi$(8,2)} &  (7,1) & 9 \\
 \cline{2-3}
&(6,2) & 10 \\
\cline{2-3}
&(5,3) & 11 \\
\cline{2-3}
&(4,4) &12 \\
\hline
\multirow{5}{*}{$\phi$(8,3)} & (6,1,1) & 11 \\
\cline{2-3}
&(5,2,1) & 12 \\
\cline{2-3}
&(4,3,1) & 13 \\
\cline{2-3}
&(4,2,2) & 14 \\
\cline{2-3}
&(3,3,2) & 15\\
\hline
\multirow{5}{*}{$\phi$(8,4)} & (5,1,1,1) & 14 \\
\cline{2-3}
&(4,2,1,1) & 15 \\
\cline{2-3}
&(3,3,1,1) & 16 \\
\cline{2-3}
&(3,2,2,1) & 17 \\
\cline{2-3}
&(2,2,2,2) & 20 \\
\hline
\multirow{3}{*}{$\phi$(8,5)} & (4,1,1,1,1) & 18 \\
\cline{2-3}
&(3,2,1,1,1) & 19 \\
\cline{2-3}
&(2,2,2,1,1) & 21 \\
\hline
\multirow{2}{*}{$\phi$(8,6)} & (3,1,1,1,1,1) & 23 \\
\cline{2-3}
&(2,2,1,1,1,1) & 24 \\
\hline
$\phi$(8,7) & (2,1,1,1,1,1,1) & 29 \\
\hline
$\phi$(8,8) & (1,1,1,1,1,1,1,1) & 36 \\
\hline
\end{tabular}
\end{center}
\caption{\small $\phi(8)$. }
\end{table}

%Algorithm \ref{genere} generates  the set of  $\phi(|V|)$ for a graph $G=(V,E)$ in the order defined above.  
%It calls  Algorithm  \ref{mot}  ($MOTIF$) to generate all motifs of $\phi(n,k)$ in the decreasing lexicographical order. 
%
%\begin{algorithm}
%\caption{GENERATE($G$) }
%\Input{ $G=(V,E)$  with $\vert V \vert=n$}
%\output {$\phi (n)$}
%	  \label{genere}
%\Begin{
%	$\phi (n):=\emptyset$ \;
%	
%	\For{$k \leftarrow 1$ downto $n$}{
%			$\phi(n,k)=\emptyset$\;
%			$ MOTIF(k,n,\emptyset,\phi(n,k))$ \;
%			$\phi(n)\leftarrow \phi(n)\cup \phi(n,k)$\;
%			
%					
%		}
%	\Return {$\phi(n$)\;}
%	
%	
%	}
%
%\end{algorithm}

% Sinon nous calculons la valeur supérieure et inférieure que peut prendre la $partie$. Dans le cas où la valeure maximale de la $partie$ et supérieure à la valeure de la partie précédente, la partie prend alors la valeur de cette dernière (ligne 6 algorithme 2). Puis un appel récursif à $MOTIF$ est effectué en faisant varier la valeur de la $partie$ courant du motif de $max$ à $inf$. Ce qui nous permet donc d'obtenir l'ensemble des motifs de $\phi(n,k)$, trié par ordre lexicographique décroissant.\\

Algorithm \ref{mot} generates the motifs of $\phi(n,k)$ in the decreasing lexicographic order, so that we can assign an index $i$ to each motif in $\phi(n,k)$. 

\begin{definition}
We denote by $p^i_k$ the $i^{th}$ motif in $\phi(n,k)$.  
%Given any two motifs $p^i_k$ and $p^j_{k'}$, $p^i_k$ precedes $p^j_{k'}$ 
%if $k<k'$ or if $k=k'$ and $p^i_k$ is lexicographically superior to  $p^j_{k'}$.
\label{ordomotif}
\end{definition}

The first motif $p^1_k$ in $\phi(n,k)$ is 

$$(n-k+1, \overbrace{1, 1, \ldots, 1}^{k-1 ~times})$$

and the last motif in $\phi(n,k)$ is

$$(\overbrace{\left\lceil \frac{n}{k}\right\rceil, \ldots, \left\lceil \frac{n}{k}\right\rceil}^{n ~mod~ k  ~times}, 
\overbrace{\left\lfloor\frac{n}{k}\right\rfloor, \ldots, \left\lfloor\frac{n}{k}\right\rfloor}^{k ~-~ n ~mod ~k  ~times})$$

Algorithm \ref{mot} works as follows. Given two positive integers $n$ and $k$ such that $k\leq n$, and a motif $p$ under construction that is initially empty, the algorithm generates all partitions of $n$ into $k$ integers in the decreasing lexicographic order. Each partition begins by the first integer that should be between $sup$ that is $n-k+1$ if $p$ is empty, $min(n-k+1, min(p))$ otherwise, and $\left\lceil \frac{n}{k}\right\rceil$, the other integers being generated by a recursive call of the algorithm. Note that the first integer should not be bigger than the minimum integer in $p$ if $p$ is not empty. The integers of each partition are appended to the end of $p$ to form a complete motif in decreasing order.

For example, in the call MOTIF(3, 8, $\emptyset$, $\emptyset$) ($p$=$\phi$=$\emptyset$), the algorithm generates each first integer 
between 6 and 3. When the first integer is 6, the algorithm calls MOTIF(2, 2, \{6\}, $\emptyset$) which generates only one partition of 2 into 2 integers (i.e. (1, 1)) appended into \{6\} to form a complete motif (6, 1, 1). When the first integer is 4, the algorithm calls MOTIF(4, 2, \{4\}, \{(6, 1, 1), (5, 3, 2)\}) that generates two partitions of 4 into 2 integers: (3, 1) and (2, 2), each partition being appended into \{4\} to form a complete motif. See Table $2$ for all results.

\begin{algorithm}
\caption{\small MOTIF($n,k,p,\phi$) }
\Input{two positive integers $n$ and $k$ such that $k \leq n$,  motif $p$ under construction, set of motifs $\phi$ under construction}
\Output{a set of motifs $\phi$ in decreasing lexicographic order}
	  \label{mot}
\Begin{
	\If{$k=1$}{ 
		$\phi\leftarrow \phi\cup \{p\cup \{n\}\}$ \;
		
	} 
	\Else{
		%{\bf if} $p$=$\emptyset$ $x\leftarrow n$ {\bf else} $x\leftarrow min(p)$\;
		\If{$p$=$\emptyset$}{
		   $sup\leftarrow n-k+1$;
		}
		\Else{
		   $sup\leftarrow min(n-k+1, min(p))$;
		}
		$inf\leftarrow \left\lceil \frac{n}{k}\right\rceil $ \;

		\ForEach{$x \leftarrow sup$ downto $inf$}{
			
			Motif($n-x, k-1,p\cup\{x\},\phi$) \;

		}
		
	}
	
	}

\end{algorithm}

\subsection{Principle of the new upper bounds}
Given a valid coloring $X$ of a graph $G$ and its associated motif $p$, we will find the number $k_{t}$ of colors such that all motifs with $k_{t}$ or more colors are dominated by $p$. Namely:

\begin{equation}
\forall q \in \psi=\bigcup\limits_{x=k_{t}}^{n}\phi(n,x), \ p \succeq q 
\end{equation}

Note that $\psi$ is not empty, because it contains trivially the only motif (1, 1, \ldots, 1) in $\phi(n, n)$ for $k_t=n$. Since $\psi$ does not contain any valid coloring better than $p$, $k_t$ is an upper bound of $s(G)$. The problem is to make $k_t$ as small as possible.

Based on this above principle, we will establish two upper bounds for $s(G)$: UB$_a$, an algebraic upper bound,  and UB$_s$, an algorithmic upper bound based on a maximum stable set of $G$.

\subsection{UB$_a$, an algebraic upper bound for $s(G)$}

UB$_a$ is based on the following two properties.

\begin{prop}
\label{1stMotif}
Let $G=(V,E)$ with $|V|$=$n$, and $k \leq n$ be an integer. The motif $p^1_k=(n-k+1, 1, 1, \ldots, 1)$ dominates all other motifs in $\phi(n,k)$.
\end{prop}

In fact, let $q\in \phi(n,k)$. Then $\forall t$ such that $1\leq t \leq k$, $\sum\limits_{x=t+1}^{k}p^1_k[x] \leq \sum\limits_{x=t+1}^{k}q[x]$, because $p^1_k[x]= 1 \leq q[x]$ when $x>1$. So, 
$\sum\limits_{x=1}^{t}p^1_k[x]=n-\sum\limits_{x=t+1}^{k}p^1_k[x] \geq n- \sum\limits_{x=t+1}^{k}q[x] = \sum\limits_{x=1}^{t}q[x]$.

\begin{prop}
\label{1stMotifs}
Let $G=(V,E)$ with $|V|$= $n$, and $k$ and $k'$ be two integers such that $1$$\le$$k$$<$$k'$$\le$$n$. The motif $p^1_k=(n-k+1, 1, 1, \ldots, 1)$ dominates
$p^1_{k'}=(n-k'+1, 1, 1, \ldots, 1)$.
\end{prop}

In fact, since $n-k+1 > n-k'+1$, $\forall t$ such that $1\leq t \leq k$, $\sum\limits_{x=1}^{t}p^1_k[x] \geq \sum\limits_{x=1}^{t}p^1_{k'}[x]$.

Property \ref{1stMotif} and Property \ref{1stMotifs} mean that the motif $p^1_{k}$=(n-k+1, 1, 1, \ldots, 1) dominates any motif in $\phi(n,k')$ such that $k\leq k'$. The sum coloring of $p^1_k$ is :

\begin{equation}
\label{p1k}
\Sigma(p^1_{k})=(n-k+1)+\sum\limits_{x=2}^k x = \frac{1}{2}k^2-\frac{1}{2}k+n
\end{equation}

So UB$_a$ is the smallest number max($k$-1, $|X|$) of colors such that

\begin{equation}
\label{eqnUBa}
\Sigma(p^1_{k}) \geq \Sigma(X)
\end{equation}

Equation \ref{eqnUBa} means that a sum coloring with $k$ or more colors cannot be better than the known valid coloring $X$ according to Property \ref{1stMotif} and Property \ref{1stMotifs}. So, the optimal solution must be with at most $k$-1 colors if $X$ uses fewer than $k$ colors, $k$ otherwise.

Using Equation \ref{p1k}, Equation \ref{eqnUBa} is transformed to 
\begin{equation}
\label{eqp1}
\frac{1}{2}k^2-\frac{1}{2}k+n -\Sigma(X) \geq 0
\end{equation}

Equation \ref{eqp1} is valid if and only if: 

\begin{equation*}
\label{equaSeuil}
k\geq  \frac{1+\sqrt{1+8\times(\Sigma(X)-n)}}{2} 
\end{equation*}

Thus:

\begin{equation}
\label{equaUB}
UB_{a}=max(\left\lceil \frac{1+\sqrt{1+8\times(\Sigma(X)-n)}}{2}\right\rceil -1, |X|)
\end{equation}

UB$_a$ is a new algebraic upper bound of $s(G)$. Note that apart from the valid coloring $X$, UB$_a$ does not consider any other structural information of $G$. We will show in the next subsection that we can obtain a better upper bound by taking into account more structural information of $G$.

\subsection{UB$_s$, an algorithmic upper bound of $s(G)$}
\label{sectionUBs}
Although MaxClique, \gcp and \mscp are all NP-hard problems, MaxClique is relatively easier to solve than \gcp and \mscp in practice. For example, the state-of-the-art exact algorithm IncMaxClique \cite{ChuInc} finds a maximum clique of any random graph of 200 vertices in few seconds, but no exact algorithm in our knowledge is able to find the chromatic number of a random graph of 200 vertices and density 0.5 in reasonable time (a random graph of density $d$ is generated by making each pair of vertices adjacent with probability $d$). So, we can find a maximum stable set of a graph $G$, which is a maximum clique in the complement graph $\overline G$, to compute an upper bound better than $UB_a$, based on Property \ref{stable:coloring}.

\begin{prop}
\label{stable:coloring}
Let $\alpha(G)$ denote the cardinality of a maximum stable set of $G$. A motif in which the first integer is bigger than $\alpha(G)$ does not correspond to any valid coloring of $G$.
\end{prop} 

We show in the sequel how to derive an upper bound of $s(G)$ called UB$_s$ by restricting ourselves to all motifs in which the first integer is smaller than or equal to $\alpha(G)$. For this purpose, we define a notion called {\em major motif} which is motivated by the following observation. Given any valid major coloring $X$, it is sometimes possible to find two color classes $X_i$ and $X_j$ ($i$$<$$j$) such that a vertex of $X_j$ can be moved into $X_i$ to obtain another valid major coloring $X'$. Clearly $\Sigma(X')$$<$$\Sigma(X)$ and the motif $p'$ associated with $X'$ dominates the motif $p$ associated with $X$. The motif $p'$ is obtained by decrementing the $j^{th}$ integer of $p$ by 1 and incrementing the $i^{th}$ integer of $p$ by 1. We call this transformation of $p$ {\em left-shifting} operation.

Note that the left-shifting operation is not always possible, because one must keep the integers of the resulting motif in non-increasing order. Moreover, all integers in the resulting motif should be positive.
Given two integers $n$ and $k$, we are interested in those motifs in $\phi(n,k)$, called {\em major motifs}, that cannot be transformed into another motif in $\phi(n,k)$ by a left-shifting operation without incrementing the first integer. 

\begin{exemple}
Consider $\phi(8,4)$ (refer to Table $2$). The major motifs are (5,1,1,1), (4,2,1,1), (3,3,1,1) and (2,2,2,2). In fact, unless the first integer is incremented, these motifs cannot be transformed into any motif in $\phi(8,4)$ by a left-shifting operation.
\end{exemple}

%We exclude the left-shifting operations incrementing the first integer of a motif, because the incrementing of the first integer may make it bigger than $\alpha(G)$, so that the obtained motif does not correspond to any valid coloring of $G$.

Let $\lambda$ denote the first integer of a motif. We have $\left\lceil \frac{n}{k}\right\rceil \leq \lambda\leq n-k+1$. Intuitively, a motif is major if it contains the maximum number ($\beta$) of integers equal to its first integer, $\lambda$. The remaining value of $n$ (i.e. $n-\beta\times\lambda$) should be partitioned into $k-\beta$ positive integers. So $\beta$ is the maximum integer satisfying $n-\beta\times\lambda \geq k-\beta$, or $\beta\leq\frac{n-k}{\lambda-1}$ after excluding the trivial motif (1, 1, \ldots, 1) and assuming $\lambda>1$. So, $\beta=\left \lfloor \frac{n-k}{\lambda - 1}\right\rfloor$.

A major motif is formally defined in Definition \ref{maj}.

\begin{definition} \label{maj}
Let $\lambda$ and $\beta$ be two integers such that $\left\lceil \frac{n}{k}\right\rceil \leq \lambda \leq n-k+1$ and $\beta=\left \lfloor \frac{n-k}{\lambda - 1}\right\rfloor$. A major motif in $\phi(n,k)$ is a motif $p^i_k$ with the following properties : 

$$
\begin{cases}

p^i_k[x]=\lambda, ~if~ 1\leq x \leq \beta ; \\
p^i_k[x]=n-\beta \times \lambda-(k-\beta-1),~if~   x=\beta +1 ; \\
p^i_k[x]=1,~if~\beta +1<  x \leq k. \\

\end{cases}
$$
\end{definition}

The interest of the major motif notion lies in the following two properties. The first one, Property \ref{majorWithSame1st}, says that a major motif in $\phi(n, k)$ dominates all motifs $\phi(n, k)$ with the same first integer and represents the best sum coloring among these motifs.

\begin{prop}
\label{majorWithSame1st}
Let $p$ and $q$ be two motifs in $\phi(n,k)$ such that $p$ is major and $p[1]=q[1]$, then $p\succeq q$.
\end{prop}
\begin{preuve} 
Let $\beta=\left \lfloor \frac{n-k}{p[1] - 1}\right\rfloor$. Since $p$ is major, we have $p[1]=p[2]=\ldots=p[\beta]=q[1]\geq q[2]\geq\ldots\geq q[\beta]$. So, $\forall t$ such that $1\leq t\leq\beta$, $\sum\limits_{x=1}^t p[x] \geq \sum\limits_{x=1}^t q[x]$. 

Moreover, $\forall t$ such that $\beta+1\leq t \leq k$, we have $p[t+1]=p[t+2]=\ldots=p[k]=1\leq q[k] \leq q[k-1] \leq \ldots \leq q[t+1]$. So,  $\sum\limits_{x=1}^t p[x]=n-\sum\limits_{x=t+1}^k p[x]\geq n-\sum\limits_{x=t+1}^k q[x]=\sum\limits_{x=1}^t q[x]$.

Therefore, $p\succeq q$. $\square$
\end{preuve}

Property \ref{2majors} compares two majors motifs in $\phi(n, k)$.

\begin{prop}
\label{2majors}
Let $p$ and $q$ be two major motifs in $\phi(n, k)$. If $p[1]>q[1]$ then $p\succeq q$.
\end{prop}
\begin{preuve}
Let $\beta_{p}=\left \lfloor \frac{n-k}{p[1] - 1}\right\rfloor$ and $\beta_{q}=\left \lfloor \frac{n-k}{q[1] - 1}\right\rfloor$. Clearly $\beta_{p}\leq\beta_{q}$. We have $p[1]=p[2]=\ldots=p[\beta_p] > q[1]=q[2]=\ldots=q[\beta_p]$. So, $\forall t$ such that $1\leq t\leq\beta_p$, $\sum\limits_{x=1}^t p[x] > \sum\limits_{x=1}^t q[x]$.

Moreover, $\forall t$ such that $\beta_p+1\leq t \leq k$, we have $p[t+1]=p[t+2]=\ldots=p[k]=1 \leq q[k] \leq q[k-1]\ldots, \leq q[t+1]$, implying  $\sum\limits_{x=1}^t p[x]=n-\sum\limits_{x=t+1}^k p[x]\geq n-\sum\limits_{x=t+1}^k q[x]=\sum\limits_{x=1}^t q[x]$. Therefore, $p\succeq q$. $\square$
\end{preuve}

Since motifs in $\phi(n,k)$ are in decreasing lexicographical order, Property \ref{majorWithSame1st} and Property \ref{2majors} imply that a major motif in $\phi(n, k)$ dominates all subsequent motifs in $\phi(n, k)$, as stated in Property \ref{dom}.

\begin{prop} 
Let $p^i_k\in \phi(n,k)$ be  a major motif and $ p^j_k \in \phi (n,k)$ such that $j>i$. Then $p^i_k\succeq p^j_k$. 
\label{dom}
\end{prop}

Observe that while a major coloring is the best sum coloring in a set of symmetric colorings, a major motif is the best motif in a set of motifs in the sense of Property \ref{dom}.
A direct consequence  is that  $ \forall p^i_k \in \phi(n,k)$, $p^1_k \succeq p^i_k $, because $p^1_k$ is major. 

The following property compares the major motif $p^1_k$ to  $p^i_{k'}$ such that $k<k'$.

\begin{prop} 
\label{prop6}
Let $k$ and $k'$ be two integers such that $k<k'$, and $\psi=\bigcup\limits_{x=k'}^{n}\{\phi(n,x)\}$, then $\forall q \in \psi$, $p^1_k\succeq q$.  
\end{prop}

\begin{preuve}
According to Property \ref{1stMotifs}, $p^1_k$ dominates $p^1_x$ $\forall x$ such that $k'\leq x \leq n$.
%because $p^1_k[1]=n-k+1>n-x+1=p^1_x[1]$. 
Since $p^1_x$ dominates all motifs in $\phi(n, x)$ according to Property \ref{dom}, $p^1_k$ dominates all motifs in $\psi$. $\square$
\end{preuve}

We now prove the most important property of this paper.

\begin{prop} 
\label{majorProp}
Let $k$ and $k'$ be two integers such that $k<k'$. Let $p^i_k$ be a major motif and $\psi=\bigcup\limits_{y=k^\prime}^{n}\{\phi(n,y)\}$. Then $\forall q \in \psi$ such that $q[1]\leq p^i_k[1]$,  $p^i_k\succeq q$. 
\end{prop}

\begin{preuve}
Let $\beta=\left \lfloor \frac{n-k}{p^i_k[1] - 1}\right\rfloor$. Since $p^i_k$ is major, we have $p^i_k[1]=p^i_k[2]=\ldots=p^i_k[\beta] \geq q[1]\geq q[2]\geq\ldots\geq q[\beta]$. So, $\forall t$ such that $1\leq t \leq \beta$, $\sum\limits_{x=1}^t p^i_k[x] \geq \sum\limits_{x=1}^t q[x]$.

Moreover, $\forall t$ such that $\beta+1\leq t \leq k$, we have $p^i_k[t+1]=p^i_k[t+2]=\ldots=p^i_k[k]=1 \leq q[k] \leq q[k-1] \leq \ldots \leq q[t+1]$ ($q$ is a sequence of non-increasing positive integers), implying  $\sum\limits_{x=1}^t p^i_k[x]=n-\sum\limits_{x=t+1}^k p^i_k[x]\geq n-\sum\limits_{x=t+1}^k q[x]=\sum\limits_{x=1}^t q[x]$. Therefore, $p^i_k\succeq q$. $\square$

\end{preuve}

Given a valid coloring $X$ of a graph $G$ with $n$ vertices and the cardinality $\alpha(G)$ of a maximum stable set of $G$, we can search for the smallest $k$ and the major motif $p^i_k$ with $p^i_k[1]=\alpha(G)$ such that $\Sigma(X) < \Sigma(p^i_k)$. Property \ref{majorProp} says that a valid coloring of $G$ with $k$ or more colors better than $X$ cannot be found. So UB$_s=k-1$. Algorithm $2$ implements UB$_s$. The function $constructMajorMotif(\lambda,k)$ constructs a major motif $p\in \phi(n,k)$  such that $p[1]=\lambda$, as defined in Definition \ref{maj}. The function $computeSumColoring(p)$ computes the sum of color weights of $p$, as shown in Equation \ref{equaMotif}.

%\begin{algorithm}[h]
%\label{UBs}
%\caption{UB$_s$($n,\alpha(G),X$) }
%\Input{ the number of vertices $n$, the cardinality of a maximum stable set $\alpha(G)$, and a valid coloring $X$}
%\Output{an upper bound of $s(G)$}
%\Begin{
%	$SUM \leftarrow 0$ \;
%	$k\leftarrow \lfloor \frac{n}{\alpha(G)} \rfloor - 1$; ~~~~~~/*initialize $k$ */\\
%	$v\leftarrow \alpha (G)$\;
%	\While{$SUM \leq \Sigma(X)$}{
%		/* Compute the sum of color weights for a major motif $p^i_k$ */\;
%		$k\leftarrow k+1$ \;	
%		\If {$p^1_k[1] < v$ }
%			{	$v\leftarrow p^1_k[1]$\;
%			}
%		/* color weight  */ \\				
%		$w\leftarrow 1$ \;
%		$SUM \leftarrow 0$ \;
%		$\beta \leftarrow \lfloor \frac{n-k}{\alpha(G) -1}\rfloor$ \;
%		/* Compute the sum coloring of $p^i_k$  */ \\		
%		\While{$w\leq \beta$}{
%		    $SUM\leftarrow SUM + w \times \alpha(G)$ \;
%		    $w \leftarrow w +1$ \:		
%		}
%		$SUM\leftarrow SUM + w \times (n-\beta\times\alpha(G)  - (k - \beta -1)) $ \;
%		\While{$w < k$}{
%		   $w\leftarrow w +1$ \;
%		   $SUM \leftarrow SUM + w$ \;
%		}
%	}
%	\Return {$k-1$\;}
%}
%\end{algorithm}

\begin{algorithm}[h]
\label{UBss}
\caption{\small CUB$_s$($n,\alpha(G),X$) }
\Input{ the number of vertices $n$, the cardinality of a maximum stable set $\alpha(G)$, and a valid coloring $X$}
\Output{an upper bound of $s(G)$}
\Begin{
	$SUM \leftarrow 0$ \;
	$k\leftarrow |X| $; ~~~~~~/*initialize $k$ */\\
	$\lambda\leftarrow \alpha (G)$\;
	\While{$SUM \leq \Sigma(X)$}{
		$k\leftarrow k+1$ \;	
		\If {$n-k+1 < \lambda$ }
			{	$\lambda\leftarrow n-k+1$\;
			}
		$p\leftarrow constructMajorMotif(\lambda,k)$\;
		$SUM\leftarrow computeSumColoring(p)$\;
	}
	\Return {$k-1$\;}
}
\end{algorithm}

The complexity of the $constructMajorMotif(\lambda,k)$ function and the complexity of the $computeSumColoring(p)$ function are both  $O(k)$. Since $k\leq n$, the complexity of Algorithm $2$ is $O(n^2)$.

\section{Empirical evaluation}

In this section, we compare our algebraic  and algorithmic  bounds (UB$_a$ and UB$_s$) for the chromatic strength of a graph $G$ with that proposed by Hajiabolhassan et al. \cite{hajia} (Property \ref{Haj}). To find a maximum stable set of $G$, we run the state-of-the-art exact MaxClique algorithm IncMaxCLQ \cite{ChuInc} to find a maximum clique in the complement graph $\overline G$. We evaluate our approach using the COLOR \cite{color023} and DIMACS \cite{dimacs2} graphs.
 
Table $3$ gives the experimental results. For each graph $G$, we denote by $k^*$ the best-known upper bound for the chromatic number ($\chi(G)$), $\Sigma^*$ the best-known upper bound for the chromatic sum ($\Sigma(G)$), $\alpha(G)$ the cardinality of a maximum stable set of $G$, Time$_s$  the runtime in seconds to find a maximum stable set in $G$, $UB_{hmt}$ the results of Hajiabolhassan et al. given by Property \ref{Haj} using $\Delta(G)$ and $k^*$, UB$_a$ the results of our algebraic bound and UB$_s$ the results of our algorithmic bound based on a maximum stable of $G$. Both UB$_a$ and UB$_s$ are computed using $\Sigma(G)^*$.

%
%\begin{landscape}
\small
\begin{longtable}[t]{|c|c|c|c|c|c|c|c|c|c|c|}

\hline
Graph  & $ |V|$ & $|E|$ & $k^*$ & $\Sigma^* $& $\Delta(G)$ &$\alpha(G)$ & Time$_s$ & $UB_{hmt}$  & UB$_a$  & UB$_s$ \\
\endhead
\hline
\multicolumn{10}{r}{{Continued on next page\ldots}}
\endfoot
\endlastfoot
\hline

anna & 138 & 493 & 11 & 276 & 71 & 80 & 0 & 41 & 17 & 14 \\
\hline
david & 87 & 406 & 11 & 237 & 82 & 36 & 0 & 47 & 17 & 15 \\
\hline
DSJC1000.1 & 1000 & 49629 & 21 & 9931 & 127 & N/A & N/A  &  74 & 134 & N/A  \\
\hline
DSJC1000.5 & 1000 & 249826 & 87 & 41603 & 551 & 15 & 408 & 319 & 285 & 187 \\
\hline
DSJC1000.9 & 1000 & 449449 & 224 & 106452 & 924 & 6 & 223 & 574 & 459 & 360 \\
\hline
DSJC125.1 & 125 & 736 & 5 & 326 & 23 & 34 & 0 & 14 & 20 & 11 \\
\hline
DSJC125.5 & 125 & 3891 & 17 & 1012 & 75 & 10 & 0 & 46 & 42 & 29 \\
\hline
DSJC125.9 & 125 & 6961 & 44 & 2503 & 120 & 4 & 0 & 82 & 69 & 58 \\
\hline
DSJC250.1 & 250 & 3218 & 8 & 996 & 38 &44 & 117 & 23 & 39 & 23 \\
\hline
DSJC250.5 & 250 & 15668 & 28 & 3306 & 147 & 12 & 0 & 88 & 78 & 53 \\
\hline
DSJC250.9 & 250 & 27897 & 72 & 8288 & 234 & 5 & 24 & 153 & 127 & 105 \\
\hline
DSJC500.1 & 500 & 12458 & 12 & 2997 & 68 & N/A &  N/A & 40 & 71 & N/A  \\
\hline
DSJC500.5 & 500 & 62624 & 48 & 11759 &  286 & 13 & 16 & 167 & 150 & 97 \\
\hline
DSJC500.9 & 500 & 112437 & 126 & 30313 & 471 & 5 & 84 & 299 & 244 & 190 \\
\hline
flat1000-50-0 & 1000 & 245000 & 50 & 39315 & 520 & 20 & 263 & 285 & 277 & 212 \\
\hline
flat1000-60-0 & 1000 & 245830 & 60 & 40648 & 524 & 17 & 296 & 292 & 282 & 200 \\
\hline
flat1000-76-0 & 1000 & 246708 & 76 & 41199 & 532  &15 & 466 & 304 & 284 & 183 \\
\hline
flat300-20-0 & 300 & 21375 & 20 & 3150 & 160 &15 & 0 & 90 & 76 & 20 \\
\hline
flat300-26-0 & 300 & 21633 & 26 & 3966 & 158 &12 & 0 & 92 & 86 & 36 \\
\hline
flat300-28-0 & 300 & 21695 & 28 & 4330 & 162 &12 & 0 & 95 & 90 & 53 \\
\hline
fpsol2.i.1 & 496 & 11654 & 65 & 3403 & 252 &307  & 0 & 159 & 76 & 75 \\
\hline
fpsol2.i.2 & 451 & 8691 & 30 & 1668 & 346 &261 & 0 & 188 & 49 & 46 \\
\hline
fpsol2.i.3 & 425 & 8688 & 30 & 1636 & 346  &238 & 0 & 188 & 49 & 46 \\
\hline
games120 & 120 & 638 & 9 & 443 & 13 & 22 & 0 & 11 & 25 & 15 \\
\hline
huck & 74 & 301 & 11 & 243 & 53 &27 & 0 & 32 & 18 & 16 \\
\hline
inithx.i.1 & 864 & 18707 & 54 & 3676 & 502 &566 & 0 & 278 & 75 & 72 \\
\hline
inithx.i.2 & 645 & 13979 & 31 & 2050 & 541 &365 &  0 & 286 & 53 & 48 \\
\hline
inithx.i.3 & 621 & 13969 & 31 & 1986 & 542 &360 & 0 & 287 & 52 & 48 \\
\hline
jean & 80 & 254 & 10 & 217 & 36 & 38 & 0 & 23 & 17 & 15 \\
\hline
le450-15a & 450 & 8168 & 15 & 2740 & 99 &75 &  45 & 57 & 68 & 53 \\
\hline
le450-15b & 450 & 8169 & 15 & 2733 &94 & 78 & 7 & 55 & 68 & 54 \\
\hline
le450-15c & 450 & 16680 & 15 & 3829 & 139 &41 & 1602 & 77 & 82 & 57 \\
\hline
le450-15d & 450 & 16750 & 15 & 3751 & 138  &41 & 2266 & 77 & 81 & 56 \\
\hline
le450-25a & 450 & 8260 & 25 & 3291 & 128  &91 & 0 & 77 & 75 & 66 \\
\hline
le450-25b & 450 & 8263 & 25 & 3492 & 111 &78 & 0 & 68 & 78 & 68 \\
\hline
le450-25c & 450 & 17343 & 25 & 4906 & 179 &47 & 24 & 102 & 94 & 79 \\
\hline
le450-25d & 450 & 17425 & 25 & 4953 & 157 &43 & 82 & 91 & 95 & 78 \\
\hline
le450-5a & 450 & 5714 & 5 & 1350 & 42  &N/A & N/A & 24 & 42 & N/A \\
\hline
le450-5b &  450 & 5734 & 5 & 1363 & 42  &N/A & N/A & 24 & 43 & N/A \\
\hline
le450-5c & 450 & 9803 & 5 & 1356 & 66 &90 & 2 & 36 & 43 & 8 \\
\hline
le450-5d & 450 & 9757 & 5 & 1350 & 68 &90 & 2 & 37 & 42 & 5 \\
\hline
miles1000 & 128 & 3216 & 42 & 1690 & 86& 8 & 0 & 64 & 56 & 48 \\
\hline
miles1500 & 128 & 5198 & 73 & 3354 & 106 &5 & 0 & 90 & 80 & 77 \\
\hline
miles250 & 128 & 387 & 8 & 325 & 16 &44 &  0 & 12 & 20 & 14 \\
\hline
miles500 & 128 & 1170 & 20 & 712 & 38&18 & 0 & 29 & 34 & 26 \\
\hline
miles750 & 128 & 2113 & 31 & 1179 & 64 &12 & 0 & 48 & 46 & 38 \\
\hline
mulsol.2 & 188 & 3885 & 31 & 1191 & 156 &90 & 0 & 94 & 45 & 44 \\
\hline
mulsol.i.1 & 197 & 3925 & 49 & 1957 &121 &100 & 0 & 85 & 59 & 59 \\
\hline
mulsol.i.3 & 184 & 3916 & 31 & 1187 & 157&86 & 0 & 94 & 45 & 44 \\
\hline
mulsol.i.4 & 185 & 3946 & 31 & 1189 &158 &86 & 0 & 95 & 45 & 44 \\
\hline
mulsol.i.5 & 186 & 3973 & 31 & 1160 & 159&88 & 0 & 95 & 44 & 43 \\
\hline
myciel3 & 11 & 20 & 4 & 21 & 5 &5 & 0 & 5 & 5 & 4 \\
\hline
myciel4 & 23 & 71 & 5 & 45 & 11 &11 & 0 & 8 & 7 & 6 \\
\hline
myciel5 & 47 & 236 & 6 & 93 & 23&23 & 0 & 15 & 10 & 8 \\
\hline
myciel6 & 95 & 755 & 7 & 189 & 47 &47 & 0 & 27 & 14 & 11 \\
\hline
myciel7 & 191 & 2360 & 8 & 381 & 95 &95 & 0 & 52 & 20 & 15 \\
\hline
queen10-10 & 100 & 1470 & 10 & 553 & 35 &10 & 0 & 23 & 30 & 12 \\
\hline
queen11-11 & 121 & 1980 & 11 & 730 & 40 &11 & 0 & 26 & 35 & 13 \\
\hline
queen12-12 & 144 & 2596 & 12 & 940 & 43  &12 & 0 & 28 & 40 & 14 \\
\hline
queen13-13 & 169 & 3328 & 13 & 1190 & 48 &13 & 0 & 31 & 45 & 16 \\
\hline
queen14-14 & 196 & 4186 & 14 & 1478 & 51 &14 & 0 & 33 & 51 & 17 \\
\hline
queen15-15 & 225 & 5180 & 15 & 1811 & 56 &15 & 0 & 36 & 56 & 19 \\
\hline
queen16-16 & 256 & 6320 & 16 & 2190 & 59 &16 & 0 & 38 & 62 & 20 \\
\hline
queen5-5 & 25 & 160 & 5 & 75 & 16 &5 & 0 & 11 & 10 & 5 \\
\hline
queen6-6 & 36 & 290 & 6 & 138 & 19 &6 &  0 & 13 & 14 & 10 \\
\hline
queen7-7 & 49 & 476 & 7 & 196 & 24 &7 & 0 & 16 & 17 & 7 \\
\hline
queen8-12 & 96 & 1368 & 12 & 624 & 32 &8 & 0 & 22 & 33 & 12 \\
\hline
queen8-8 & 64 & 728 & 9 & 291 & 27 &8 &  0 & 18 & 21 & 10 \\
\hline
queen9-9 & 81 & 1056 & 10 & 409 & 32 &9 & 0 & 21 & 26 & 11 \\
\hline
school1 & 385 & 19095 & 14 & 2674 & 282 &41 & 2 & 148 & 68 & 45 \\
\hline
school1-nsh & 352 & 14612 & 14 & 2392 & 232 &39 & 0 & 123 & 64 & 43 \\
\hline
zeroin.i.1 & 211 & 4100 & 49 & 1822 & 111 &120 & 0 & 80 & 57 & 56 \\
\hline
zeroin.i.2 & 211 & 3541 & 30 & 1004 & 140 &127 & 0 & 85 & 40 & 39 \\
\hline
zeroin.i.3 & 206 & 3540 & 30 & 998 & 140 &123 & 0 & 85 & 40 & 39 \\
\hline
\caption{\small Comparison of our new algebraic bound (UB$_a$) and algorithmic bound (UB$_s$) with the bound of Hajiabolhassan et al. \cite{hajia} (UB$_{hmt}$). }
\label{tab1}
\end{longtable}

The results in Table $3$ shows that our algebraic bound (UB$_a$) is already better than the results  of Hajiabolhassan et al. given by Property \ref{Haj} (UB$_{hmt}$), except for some graphs with low degree ($le450$, $queen$). UB$_a$ gives a better lower bound for 42 instances among 74. However, when a maximum stable set can be found in reasonable time using IncMaxCLQ, our algorithmic bound (UB$_s$) is significantly better than UB$_{hmt}$ in general. UB$_s$ gives a better lower bound for 66 instances among 74. For example, while UB$_{hmt}$ for the three graphs $inithx$ is 278, 286 and 287 respectively, UB$_s$ is 72, 48 and 48 respectively (UB$_a$ is 75, 53 and 52 respectively). Another example is the graph {\em le450-5d} for which UB$_{hmt}=37$, while UB$_s=5$. These results show the performance of our approach for reducing the number of colors to be considered when solving $MSCP$.

\section{Conclusion}

In this paper we focused on one component of the Minimum Sum Coloring Problem ($MSCP$), the chromatic strength $s(G)$ of the graph $G$. We have proposed two new upper bounds of $s(G)$, called UB$_a$ and UB$_s$ respectively. UB$_a$ and UB$_s$ both use a known valid coloring $X$ of the graph and explore a set of motifs representing an abstraction of all possible colorings of the graph. UB$_a$ is obtained by identifying the number of colors from which a coloring better than $X$ cannot be obtained.

Apart from $X$, UB$_a$ does not exploit any other structure property of the graph. UB$_s$ is a more established upper bound. In order to determine UB$_s$, we introduced a
notion called major motif that exploits the dominance relation on the set of motifs. Indeed, such a motif represents the best sum coloring among all motifs with the same or more number of colors and the same or smaller first integer.

Computing UB$_s$ consists in identifying a major motif whose the first integer is the cardinality of a maximum stable set of the graph, and whose the sum coloring is greater than the sum of $X$. The maximum stable set is computed using the exact MaxClique algorithm IncMaxCLQ. Thus, we exclude the colorings that cannot be valid for the graph and the colorings that cannot be better than $X$. UB$_s$ is derived from the remaining colorings thanks to our algorithm CUB$_s$.

We evaluated UB$_a$ and UB$_s$ on DIMACS and COLOR graphs. The experimental results show that UB$_a$ is already better than the previous bounds except for some graphs with low degree. The algorithmic upper bound UB$_s$, based on the major motif notion and a maximum stable set, outperforms generally all others bounds allowing to reduce substantially the search space when solving $MSCP$.

In the future, we plan to integrate more structural properties of a graph to further improve UB$_s$, and to develop efficient algorithms to solve $MSCP$ based on UB$_s$.

\paragraph*{Acknowledgements}
\small{ This work is supported by the Ministry of Higher Education and Research, of the French state.}

\section{References}

\bibliography{journal}
\bibliographystyle{plain}

\end{document}